# Google Scholar, Web of Science, and Scopus: a systematic comparison of citations in 252 subject categories

Alberto Martín-Martín[1] , Enrique Orduna-Malea[2] , Mike Thelwall[3] , Emilio Delgado López-Cózar[1]

Version 1.6
March 12, 2019

## Abstract

Despite citation counts from Google Scholar (GS), Web of Science (WoS), and Scopus being widely consulted by researchers and sometimes used in research evaluations, there is no recent or systematic evidence about the differences between them. In response, this paper investigates 2,448,055 citations to 2,299 English-language highly-cited documents from 252 GS subject categories published in 2006, comparing GS, the WoS Core Collection, and Scopus. GS consistently found the largest percentage of citations across all areas (93%-96%), far ahead of Scopus (35%-77%) and WoS (27%-73%). GS found nearly all the WoS (95%) and Scopus (92%) citations. Most citations found only by GS were from non-journal sources (48%-65%), including theses, books, conference papers, and unpublished materials. Many were non-English (19%-38%), and they tended to be much less cited than citing sources that were also in Scopus or WoS. Despite the many unique GS citing sources, Spearman correlations between citation counts in GS and WoS or Scopus are high (0.78-0.99). They are lower in the Humanities, and lower between GS and WoS than between GS and Scopus. The results suggest that in all areas GS citation data is essentially a superset of WoS and Scopus, with substantial extra coverage.

## Acknowledgements

Alberto Martín-Martín is funded for a four-year doctoral fellowship (FPU2013/05863) granted by the Ministerio de Educación, Cultura, y Deportes (Spain). An international mobility grant from Universidad de Granada and CEI BioTic Granada funded a research stay at the University of Wolverhampton.

---

[1] Facultad de Comunicación y Documentación, Universidad de Granada, Granada, Spain.
[2] Universitat Politècnica de València, Valencia, Spain.
[3] Statistical Cybermetrics Research Group, School of Mathematics and Computer Science, University of Wolverhampton, Wolverhampton, UK.

✉ Alberto Martín-Martín
   albertomartin@ugr.es

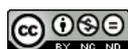 

# 1. Introduction

The launch of Google Scholar (GS) in November of 2004 brought the simplicity of Google searches to the academic environment, and revolutionized the way researchers and the public searched, found, and accessed academic information. Until that point, the coverage of academic databases depended on lists of selected sources (usually scientific journals). In contrast, and using automated methods, Google Scholar crawled the web and indexed any document with a seemingly academic structure. This inclusive approach gave GS potentially more comprehensive coverage of the scientific and scholarly literature compared to the two major existing multidisciplinary databases with selective journal-based inclusion policies, the Web of Science (WoS) and Scopus (Orduna-Malea, Ayllón, Martín-Martín, & Delgado López-Cózar, 2015).

Although citation data in Google Scholar was originally intended to be a means of identifying the most relevant documents for a given query, it could also be used for formal or informal research evaluations. The availability of free citation data in Google Scholar, together with the free software *Publish or Perish* (Harzing, 2007) to gather it made citation analysis possible without a citation database subscription (Harzing & van der Wal, 2008). Nevertheless, GS has not enabled bulk access to its data, reportedly because their agreements with publishers preclude it (Van Noorden, 2014). Thus, third-party web-scraping software is currently the only practical way to extract more data from GS than permitted by Publish or Perish.

Despite its known errors and limitations, which are consequence of its automated approach to document indexing (Delgado López-Cózar, Robinson-García, & Torres-Salinas, 2014; Jacsó, 2010), GS has been shown to be reliable and to have good coverage of disciplines and languages, especially in the Humanities and Social Sciences, where WoS and Scopus are known to be weak (Chavarro, Ràfols, & Tang, 2018; Mongeon & Paul-Hus, 2016; van Leeuwen, Moed, Tijssen, Visser, & Van Raan, 2001). Analyses of the coverage of GS, WoS, and Scopus across disciplines have compared the numbers of publications indexed or their average citation counts for samples of documents, authors, or journals, finding that GS consistently returned higher numbers of publications and citations (Harzing, 2013; Harzing & Alakangas, 2016; Mingers & Lipitakis, 2010; Prins, Costas, van Leeuwen, & Wouters, 2016). Citation counts from a range of different sources have been shown to correlate positively with GS citation counts at various levels of aggregation (Amara & Landry, 2012; De Groote & Raszewski, 2012; Delgado López-Cózar, Orduna-Malea, & Martín-Martín, 2018; Kousha & Thelwall, 2007; Martín-Martín, Orduna-Malea, & Delgado López-Cózar, 2018; Meho & Yang, 2007; Minasny, Hartemink, McBratney, & Jang, 2013; Moed, Bar-Ilan, & Halevi, 2016; Pauly & Stergiou, 2005; Rahimi & Chandrakumar, 2014; Wildgaard, 2015). See the supplementary materials [4], Delgado López-Cózar et al. (2018), Orduña-Malea, Martín-Martín, Ayllón, & Delgado López-Cózar (2016), and Halevi, Moed, & Bar-Ilan (2017) for discussions of the wider strengths and weaknesses of GS.

A key issue is the ability of GS, WoS, and Scopus to find citations to documents, and the extent to which they index citations that the others cannot find. The results of prior studies are confusing, however, because they have examined different small (with one exception) sets of articles. A summary of the results found in these previous studies is presented in Table 1. For example, the number of citations that are unique to GS varies between 13% and 67%, with the differences probably being due to the study year or the document types or disciplines covered. The only multidisciplinary study (Moed et al., 2016) checked articles in 12 journals from 6 subject areas, which is still a limited set.

---

[4] Supplementary materials available from https://dx.doi.org/10.31235/osf.io/pqr53

Table 1. Results of studies that analysed unique and overlapping citations in GS, WoS, and Scopus

| Study | Sample | N citations | % only GS | % only WoS | % only Scopus | % only GS & WoS | % only GS & Scopus | % only WoS & Scopus | % GS & WoS & Scopus | % GS (all cit.) | % WoS (all cit.) | % Scopus (all cit.) | % WoS cit. in GS | % Scopus cit. in GS |
|---|---|---|---|---|---|---|---|---|---|---|---|---|---|---|
| Bakkalbasi, Bauer, Glover, & Wang (2006) | 50 journal articles covered in JCR Oncology | 614 | 13 | 7 | 12 | 4 | 5 | 28 | 31 | 53 | 70 | 76 | 215/431 = **50%** | 220/469 = **47%** |
| | 50 journal articles covered in JCR Physics, Cond. Matter | 296 | 17 | 20 | 8 | 9 | 3 | 22 | 21 | 50 | 72 | 54 | 84/212 = **40%** | 72/162 = **44%** |
| Yang & Meho (2007) | Scientific production of two Library & Information Science (LIST) researchers | 385 | 10 | 23 | 6 | 10 | 7 | 18 | 25 | 52 | 77 | 57 | 137/295 = **46%** | 124/218 = **57%** |
| Meho & Yang (2007) | 1,457 articles published by 25 LIS researchers | 5,285 | 48 | Only (WoS or Scopus): 21 | | GS-(WoS or Scopus): 31 | | NA | NA | 79 | 38 | 44 | % (WoS or Scopus) cit. in GS 1,629/2,733 = **60%** | |
| Kousha & Thelwall (2008) | 262 WoS-covered Biology journal articles | 1,554 | 17 | 28 | NA | 55 | NA | | | 72 | 83 | NA | 847/1288 = **66%** | NA |
| | 276 WoS-covered Chemistry journal articles | 729 | 8 | 62 | | 30 | | | | 38 | 92 | | 218/668 = **33%** | |
| | 262 WoS-covered Physics journal articles | 1,734 | 36 | 24 | | 40 | | | | 76 | 64 | | 690/1111 = **62%** | |
| | 82 WoS-covered Computing journal articles | 3,369 | 67 | 14 | | 19 | | | | 86 | 33 | | 632/1117 = **57%** | |
| | Total WoS-covered journal articles (882) | 7,386 | 43 | 24 | | 32 | | | | 76 | 57 | | 2387/4184 = **57%** | |
| Jacimovic, Petrovic, & Zivkovic (2010) | 158 articles published in Serbian Dental Journal | 249 | 58 | 4 | 6 | 1 | 2 | 15 | 15 | 76 | 34 | 39 | 39/85 = **46%** | 43/94 = **46%** |
| Bar-Ilan (2010) | Book "Introduction to Informetrics" by L. Egghe and R. Rousseau | 397 | 27 | 12 | 2 | 6 | 5 | 9 | 39 | 77 | 66 | 55 | 177/259 = **68%** | 174/218 = **80%** |
| Lasda Bergman (2012) | 5 top journals in the field of Social Work | 4,308 | 44 | 5 | 8 | 2 | 8 | 12 | 22 | 76 | 41 | 50 | 1042/1741 = **60%** | 1285/2126 = **60%** |
| de Winter, Zadpoor, & Dodou (2014) | Garfield, E. (1955). Citation indexes for science. *Science*, 122(3159), 108-111. | 1,309 | 33 | 41 | NA | 35 | NA | | | 68 | 76 | NA | 453/606 = **75%** | NA |
| Rahimi & Chandrakumar (2014) | 2,082 WoS-covered articles in General and Internal Medicine | 62,900 | 29 | 10 | 11 | 2 | 9 | 8 | 31 | 71 | 51 | 59 | 20532/31778 = **65%** | 25180/37272 = **68%** |
| Moed, Bar-Ilan, & Halevi (2016) | Articles published in 12 journals from 6 subject areas | 6,941 | 47 | NA | 6 | NA | 47 | NA | NA | 94 | NA | 53 | NA | 3246/3651 = **89%** |

NA = not analysed in the study
Cells with more intense background color represent higher percentages of citations within the same sample of documents.

The fields previously compared for citation sources (Table 1) are Library and Information Science (5 out of 10 articles analyse case studies about LIS documents/journals/researchers), Medicine (3 papers, analysing oncology, general medicine, and dentistry), Physics (2 articles: general and condensed matter), Chemistry (2 articles: general and inorganic), Computer Science (2 articles: general, and computational linguistics), Biology (2 articles: general, and virology), Social Work, Political Science, and Chinese Studies (1 article each). From this list it is clear that most academic fields have not been analysed for Google Scholar coverage. The studies used small samples of documents and citations (9 out of 10 papers analysed less than 10,000 citations), probably because of the difficulty of extracting data from GS, caused by the lack of a public API (Else, 2018; Van Noorden, 2014). Moreover, the most recent data in these studies was collected in 2015 (three years before the current study), and the oldest data is from 2005 (13 years ago).

Given the limited nature of all prior studies of citing sources for GS and the need to update all previous research, a comprehensive analysis of citation sources in GS, WoS, and Scopus across all subject areas is needed. This information is important for those deciding whether to use GS citation counts for informal or formal research evaluations. The following research questions drive this investigation.

RQ1. How much overlap is there between GS, WoS, and Scopus in the citations that they find to academic documents and does this vary by subject?

RQ2. Do the citing documents that are only found by GS have a different type to non-unique GS citations, and does this vary by subject?

RQ3. How similar are citation counts in GS to those found in WoS and Scopus, at the level of subjects?

## 2. Methods

The sample used for this study is taken from GS's *Classic Papers* product (GSCP)[5]. The 2017 edition of GSCP lists 2,515 highly-cited documents written in English and published in 2006[6]. These documents were classified by GS into 252 subject categories within 8 broad subject areas. Background about GSCP can be found in Orduna-Malea, Martín-Martín, & Delgado López-Cózar (2018) and Martín-Martín, Orduna-Malea, & Delgado López-Cózar (2018). This gives a large sample of highly cited documents classified by subject. This is not a random sample of academic publications because there is no complete list of these. There is also not a complete list of documents in GS.

The GSCP sample is suitable because it covers all subject areas and, because the articles are classified, allows analyses by subject categories. GSCP and Google Scholar Metrics[7] (GSM) are the only products where GS provides a subject categorization. Taking a sample from one of the three sources to be compared (GS, Scopus, WoS) is not ideal because it is likely to bias the results in favour of GS. Nevertheless, the inclusion of 252 categories minimizes the chance of bias due to a subject area that is not well covered by GS. GS is also a better source than WoS or Scopus because of its more comprehensive coverage, as found by most prior studies.

### 2.1. Extraction of data from Google Scholar

The citations to each of the 2,515 GSCP documents were extracted from GS, WoS, and Scopus between April 22nd and May 6th, 2018. A custom script scraped all the relevant information from GS SERPs (Search Engine Results Pages) (Figure 1). Searches were submitted from

---

[5] https://scholar.google.com/citations?view_op=list_classic_articles&hl=en&by=2006
[6] https://osf.io/5zmk7/
[7] https://scholar.google.com/citations?view_op=top_venues&hl=en

Universidad de Granada IP addresses to access the additional information displayed in GS for WoS subscribers (Clarivate Analytics, 2015). CAPTCHAs were solved manually when GS requested them. This process found 2,415,072 citations in Google Scholar[8] to the 2,515 highly-cited documents. The number of citations is reduced to 2,301,997 for the 2,299 highly-cited documents also covered by WoS and Scopus.

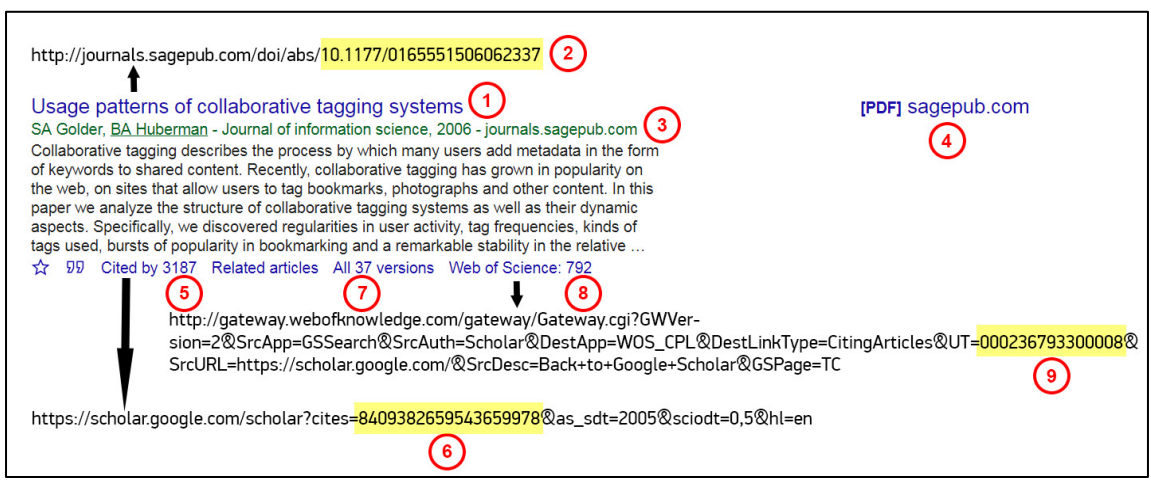

**Figure 1. Metadata extracted from Google Scholar**

1. Title of the document.
2. URL embedded in title of the document. The DOI of the document is sometimes embedded in this URL (depending on the host)
3. Authors, publication venue, publication year, and publisher or web domain that hosts the document.
4. URL to the freely accessible full text of the document, when available.
5. Times cited according to GS.
6. URL pointing to list of citing documents according to GS. GS's internal ID for the document is embedded in this URL.
7. Number of versions of the document found by GS.
8. Times Cited according to WoS (when the document is also covered by WoS).
9. URL pointing to list of citing documents in WoS. WoS's internal ID (UT) for the document is embedded in this URL.

The data was processed to clean and enrich the limited metadata available in GS, as follows.

- DOI were detected for as many citing documents as possible. The following techniques were used, retrieving 1,501,178 DOIs (62%).
    - Extracted from URLs for publishers like Wiley, Springer, and SAGE which embed the DOI in the article's landing page URL (Figure 1, #2).
    - Looked up with public APIs offered by the publishers (Elsevier, IEEE) or CrossRef[9] (using the *alternative-id* filter option), when the publisher landing page contained publisher document ID.
    - Extracted from a HTML Meta tag in the webpage from which Google Scholar extracted the document's metadata.
- Metadata was obtained from CrossRef and DataCite APIs when a DOI was available or otherwise from HTML Meta tags present in the website hosting the citation, when possible.

## 2.2. Extraction of data from Web of Science and Scopus

Each of the 2,515 highly-cited documents in GSCP was searched for in the WoS (Core Collection) web interface. The list of citations to each document was extracted (in batches of up to 500 records per download). The exported files were consolidated into a single table using a set of R functions developed for this purpose (Martín-Martín & Delgado López-Cózar, 2016). Although R has built-in functions and additional libraries to read tabulated data, none of them seemed to work with data exported from WoS. A total of 1,270,225 WoS records were collected[10]. At the time of data collection FECYT[11], the Spanish organization that manages the national subscription to Clarivate Analytics' services, had not subscribed to the Emerging Sources Citation Index (ESCI) Backfile for documents published between 2005 and 2014 (Clarivate Analytics, 2017), and so the results exclude this source.

---

[8] https://osf.io/qg8kb/
[9] https://www.crossref.org/services/metadata-delivery/rest-api/

Each of the 2,515 highly-cited documents in GSCP were also searched for in the Scopus web interface. This has a limit of 2,000 records when exporting citations. When a highly-cited document had more than 2,000 citations, these could still be extracted using the alternative email service, which allows the extraction of up to 20,000 citation records in one go. A total 1,515,436 Scopus records were collected[12].

Most of the highly-cited documents (2,299 out of 2,515) were covered by all three databases, and the citations to these 2,299 documents are analysed here.

## 2.3. Identification of document types and languages of citing documents

Unlike WoS and Scopus, GS does not provide metadata on the document type and the language of the documents that it covers. The metadata extracted from CrossRef's API and HTML Meta tags of the hosting website gave this information for 83% of the citing documents. Adding metadata from WoS and Scopus increased this percentage to 85%. The following categories were used.

- Journal publication: article, review, letter, editorial…
- Conference paper: paper presented at conference, symposium, workshop, society meeting…
- Book or book chapter: scientific/scholarly monograph
- Thesis or dissertation: document presented by student to fulfill the requirements of a doctoral, masters', or bachelor's degree
- Other not-formally-published scientific/scholarly paper: working paper, discussion paper, other paper for which no formal publication venue could be found.
- Other: report, patent, presentation slides, syllabus, educational materials, errata…
- Unknown: document for which no document type could be identified

To identify the distribution of document types in the 15% for which metadata was not available, eight random samples of 500 citing documents with an unknown document type were selected, one for each of the broad subject categories in which GSCP are classified. The document types of these 4,000 citing documents were manually identified by accessing and perusing the full text of the documents (when possible) or the available metadata. The proportion of document types found in these random samples were applied as a correction factor to the percentage of citations with an unknown document type in each broad subject area. For example, in the Social Sciences, 33.5% of the citing documents were classified as journal articles using the available metadata, but 20% of all citing documents could not be classified with the available metadata. A random sample of documents from that unknown 20% were selected and analyzed manually, finding that 27.6% of the items in the random sample were journal articles. Therefore, the total percentage of journal articles in Social Sciences was 33.5% + (27.6% of 20% = 5.5%) = 39%.

The language of 98% of the citing documents was identified by combining data from three sources (in the order of preference shown below).

---

[10] https://osf.io/6c7ta/
[11] https://www.fecyt.es/
[12] https://osf.io/n6k9w/

1. Metadata in CrossRef and HTML Meta tags.
2. Metadata in WoS (Scopus did not provide document language information).
3. Google's Compact Language Detector 2[13] applied to the document title.

For RQ1, the citations extracted from GS, WoS and Scopus were matched as follows. Three pairwise matching processes were carried out: GS–Scopus; GS–WoS; and Scopus–WoS.

1. For each pair of databases *A* and *B*, and a highly-cited document from GSCP *X*, all citing documents with a DOI that cite *X* according to *A* where matched to all citing documents with a DOI that cite *X* according to *B*.
2. For each of the unmatched documents citing *X* in *A* and *B*, a further comparison was carried out. The title of each unmatched document citing *X* in *A* was compared to the titles of all the unmatched document citing *X* in *B*, using the restricted Damerau-Levenshtein distance (optimal string alignment) (Damerau, 1964; Levenshtein, 1966). The pair of citing documents which returned the highest title similarity (1 is perfect similarity) was selected as potential matches. This match was considered successful if either of the following conservative heuristics was met.
   - The title similarity was at least 0.8, and the citing document title was at least 30 characters long (to avoid matches between titles like "Introduction").
   - The title similarity was at least 0.7, and the first author of the citing document was the same in *A* and *B*.

For RQ2, the document types, languages, and citation counts of the citing documents in our sample (see Figure 2) were aggregated or averaged by GSCP broad subject areas, differentiating between unique GS citations and overlapping citations.

For RQ3, Spearman correlation coefficients were calculated for the citation counts of the citing documents in our sample (GS-WoS, and GS-Scopus), by subject category. Correlation coefficients are considered useful in high-level exploratory analyses to check whether different indicators reflect the same underlying causes (Sud & Thelwall, 2014). In this case, however, the goal is to find out whether the same indicator, based on different data sources, provides similar relative values. Spearman correlations were used because it is well-known that the distributions of citation counts and other impact-related metrics are highly skewed (De Solla Price, 1976). For the GS-WoS comparison, WoS subject categories and (for an additional check) the NOWT classification (Tijssen et al., 2010) were used. For the GS-Scopus comparison, the ASJC (All Science Journal Classification) available in the Scopus source list (Elsevier, 2018) was used.

To carry out all these processes, the R programming language (R Core Team, 2014), and several R packages and custom functions were used (Dowle et al., 2018; Larsson et al., 2018; Martín-Martín & Delgado López-Cózar, 2016; Ooms & Sites, 2018; van der Loo, van der Laan, R Core Team, Logan, & Muir, 2018; Walker & Braglia, 2018; Wickham, 2016). The resulting data files are openly available[14].

---

[13] https://github.com/CLD2Owners/cld2
[14] https://osf.io/gnb72/

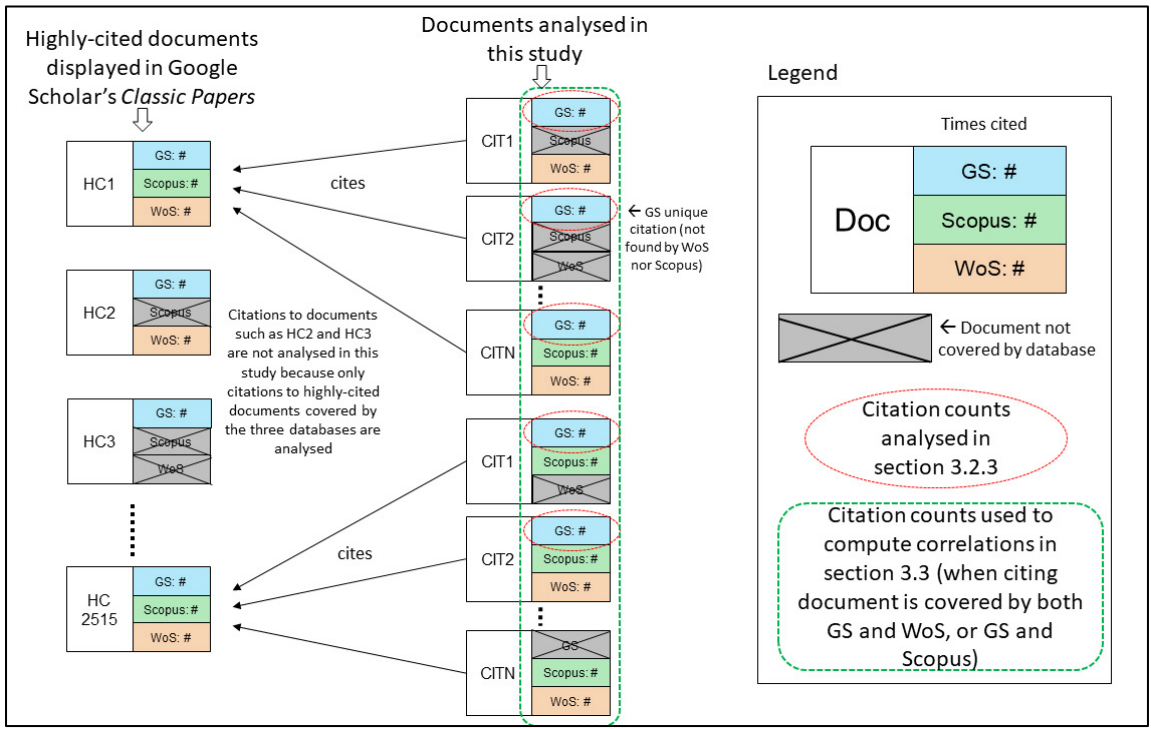

**Figure 2. Visual representation of the documents and citation counts analysed in this study**

# 3. Results

## 3.1. RQ1: Citing source overlap

Overall, 46.9% of all citations were found by the three databases (Figure 3). GS found the most citations, including most of the citations found by WoS and Scopus. In contrast, only 6% of all citations were found by WoS and/or Scopus, and not by GS. An additional 10.2% of all citations were found by both GS and Scopus (7.7%), or GS and WoS (2.5%). Over a third (36.9%) of all citations were only found by GS.

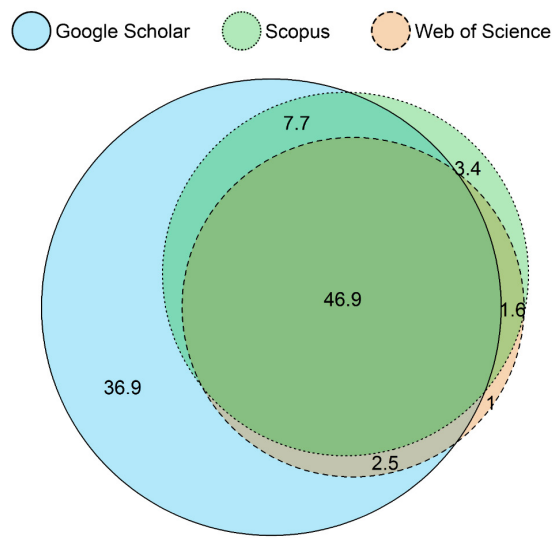

**Figure 3. Percentage of unique and overlapping citations in google scholar, Scopus, and Web of Science. n = 2,448,055 citations from all subject areas**

When citations are disaggregated by the broad subject area in which the cited document was classified according to GSCP, important differences emerge (Figure 4). In *Humanities, Literature & Arts*, *Social Sciences*, and *Business, Economics & Management* the proportion of unique GS citations is well over 50% of all citations, surpassing 60% in the case of *Business, Economics & Management*. In these categories the proportion of citations found by all three databases ranges from 21.4% (*Humanities, Literature & Arts*) to 29.8% (*Social Sciences*). On the other hand, in *Engineering & Computer Science*, *Physics & Mathematics*, *Health & Medical Sciences*, *Life Sciences & Earth Sciences*, and *Chemical & Material Sciences*, the proportion of unique GS citations is much lower (20.3% - 34.3%), and the overlap is higher: percentages of citations found by all three databases range from 46.8% (*Engineering & Computer Science*) to 67.7% (*Chemical & Material Sciences*).

For the 252 specific subject categories (data and figures for each category are available in the supplementary materials [15]), there are more extreme differences (Figure 5). The highest percentages of unique citations in GS (over 70% of all citations) are found in *Educational Administration*[16], *Foreign Language Learning*[17], *Chinese Studies & History*[18], and *Finance*[19]. On the other hand, the highest percentages of overlap in the three databases (over 70% of all citations) are found in *Crystallography & Structural Chemistry*[20], *Molecular Modeling*[21], *Polymers & Plastics*[22], and *Chemical Kinetics & Catalysis*[23].

---

[15] https://osf.io/t3sxh/

[16] https://osf.io/xfepy/

[17] https://osf.io/wk6se/

[18] https://osf.io/q8k3u/

[19] https://osf.io/56azc/

[20] https://osf.io/ysg2j/

[21] https://osf.io/cq8j6/

[22] https://osf.io/4jwta/

[23] https://osf.io/9hmf3/

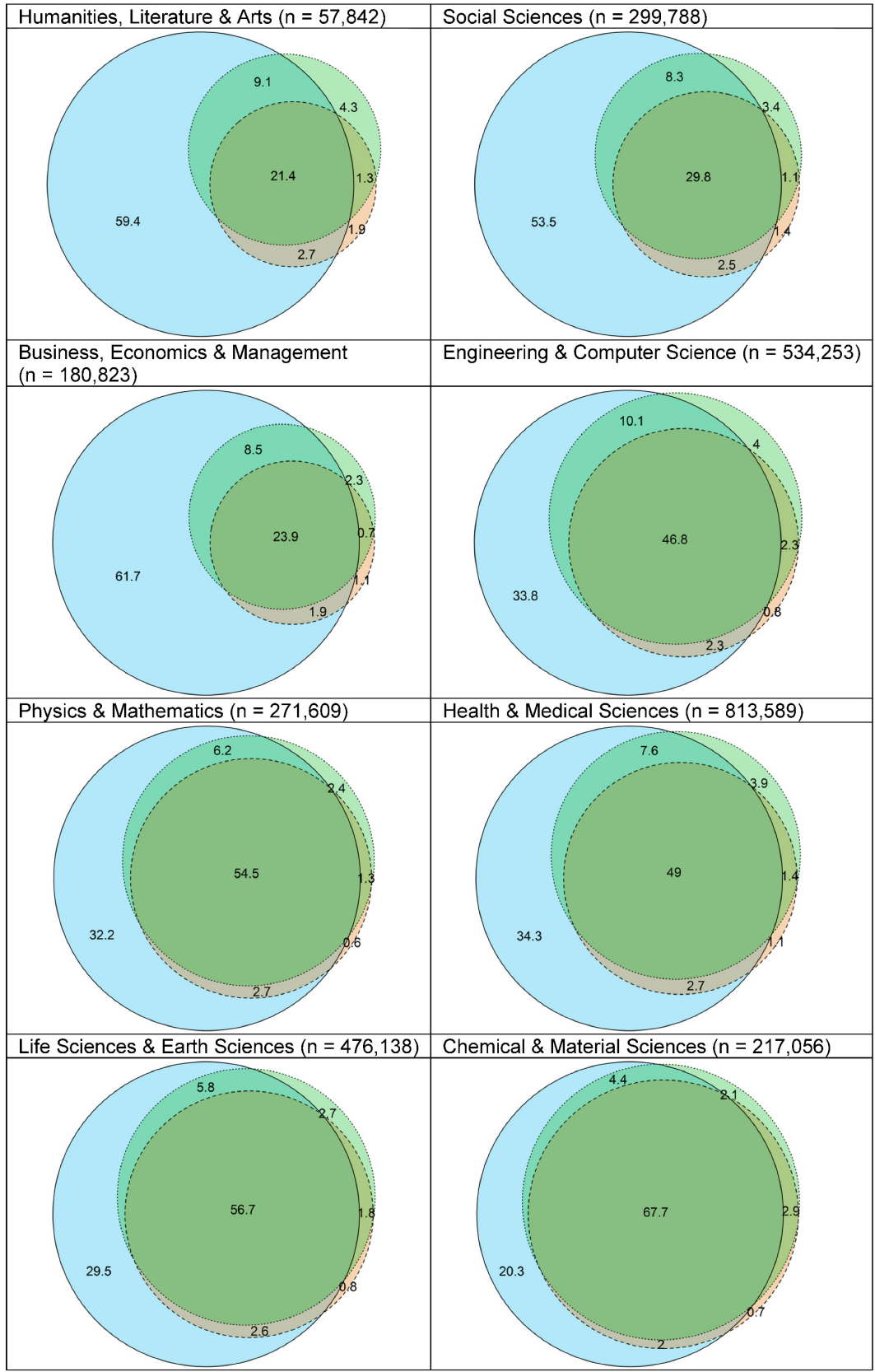

Figure 4. Percentage of unique and overlapping citations in Google Scholar, Scopus, and Web of Science, by broad subject area of cited documents

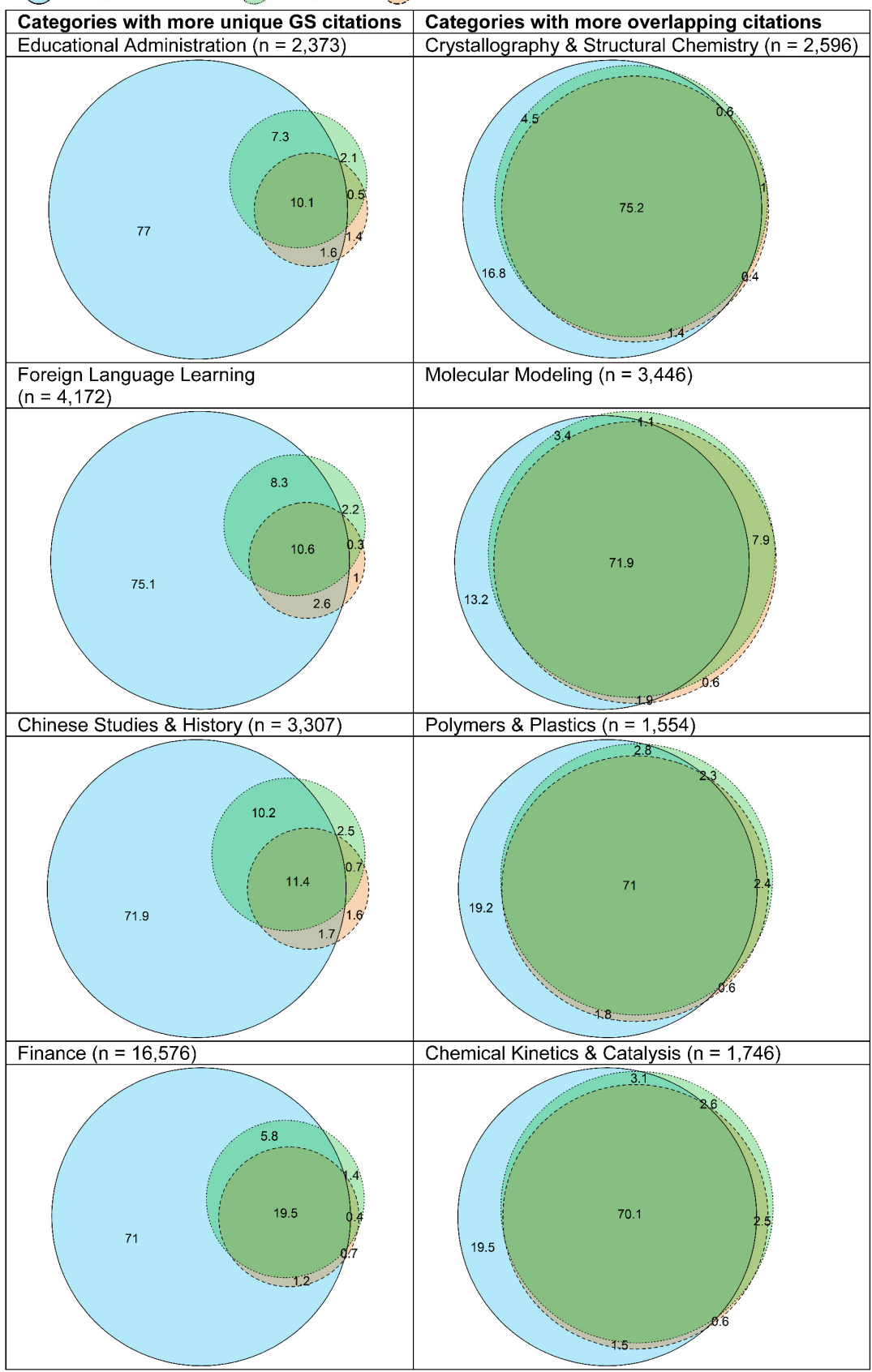

Figure 5. Categories with many unique citations or many overlapping citations

Overall, GS found 94% of all citations (93%-96% depending on the area), while WoS found 52% (ranging from 27% in *Humanities, Literature & Arts*, to 73% in *Chemical & Material Sciences*), and Scopus 60% (from 35% in *Business, Economics & Management*, to 77% in *Chemical & Material Sciences*). Additionally, GS found 95% of the citations that WoS found (88%-97% depending on the area), and 92% of the citations that Scopus found (84-94%) (Table 2). The data also shows that Scopus found 93% of the citations that Web of Science found (83-96% depending on the area).

Table 2. Percentage of citations in Google Scholar, Web of Science, and Scopus, relative to all citations, and relative to citations found by other databases

|  | % GS (all cit.) | % WoS (all cit.) | % Scopus (all cit.) | % WoS cit. in GS | % Scopus cit. in GS | % WoS cit. in Scopus |
|---|---|---|---|---|---|---|
| Overall | 94 | 52 | 60 | 95 | 92 | 93 |
| Humanities, Literature & Arts | **93** | **27** | 36 | **88** | **84** | **83** |
| Social Sciences | 94 | 35 | 43 | 93 | 89 | 89 |
| Business, Economics & Management | **96** | 28 | **35** | 93 | 92 | 89 |
| Engineering & Computer Science | **93** | 52 | 63 | 94 | 90 | 94 |
| Physics & Mathematics | 96 | 59 | 64 | **97** | **94** | 94 |
| Health & Medial Sciences | 94 | 54 | 62 | 95 | 91 | 93 |
| Life Sciences & Earth Sciences | 95 | 62 | 67 | 96 | 93 | 95 |
| Chemical & Material Sciences | 94 | **73** | **77** | 95 | **94** | **96** |

The results for the 252 specific subject categories (available in the supplementary materials[24]) show that GS covers at least 90% of all citations in 233 out of 252 categories, the lowest value being 77% in *Visual Arts*[25], and the highest values around 98% in *Crystallography & Structural Chemistry*[26], *Evolutionary Biology*[27], *Quantum Mechanics*[28], and *Astronomy & Astrophysics*[29]. Relative to the coverage of WoS and Scopus, GS finds at least 90% of the citations that WoS and Scopus find in 221 and 164 categories, respectively, the lowest values belonging to the Humanities, such as *Film*[30], *Visual Arts*[31], and *History*[32] (56%-68%).

### 3.2. RQ2. Unique and non-unique citations
### 3.2.1. Document types

The distribution of document types of unique GS citations greatly differs from that of citations that were also found by WoS and/or Scopus. This is true across all eight broad subject categories (Figure 6). Among non-unique citations, the most common document type by far is the journal publication (from 71% in *Engineering & Computer Science*, to 94% in *Chemical & Material Sciences*). The other document types present among non-unique citations are books / book chapters and conference papers, with levels varying by subject area. Among unique GS citations, however, there is more document type diversity (including many never indexed by WoS or Scopus). Although journal publications are still the single most frequent document type, other document types comprise over 50% in all subject areas except *Health & Medical Sciences* (48%). The most frequent non-journal document type is the thesis or dissertation (22% in *Business,*

---

[24] https://osf.io/t3sxh/

[25] https://osf.io/7ea63/

[26] https://osf.io/ysg2j/

[27] https://osf.io/javkb/

[28] https://osf.io/cr3k2/

[29] https://osf.io/wmn8c/

[30] https://osf.io/7dkm3/

[31] https://osf.io/7ea63/

[32] https://osf.io/fgrp4/

*Economics & Management* – 37% in *Chemical & Material Sciences*), followed by books and book chapters (especially in *Humanities, Literature & Arts* and *Social Sciences*). This trend is different in *Engineering & Computer Science*, where conference papers are more common than books, and in *Business, Economics & Management* and *Physics & Mathematics*, where unpublished scholarly papers (such as working papers and preprints) are also more frequently used than books for scientific communication.

Considering the 252 specific subject categories[33], the percentage of known document types other than journal articles in the unique GS citations ranges from approx. 10% in Nonlinear Science, Heart & Thoracic Surgery, Natural Medicines & Medicinal Plants, and Oral & Maxillofacial Surgery, to over 55% in Special Education, and Computer Hardware & Design. However, unlike in the analysis by broad subject categories, a correction factor has not been applied (because no random samples were selected and analysed at this level), and therefore the document types of a large percentage of the citations are unknown (from approx. 20% in Special Education, and Ethnic & Cultural Studies, to over 50% in Quantum Mechanics, Geometry, and Algebra).

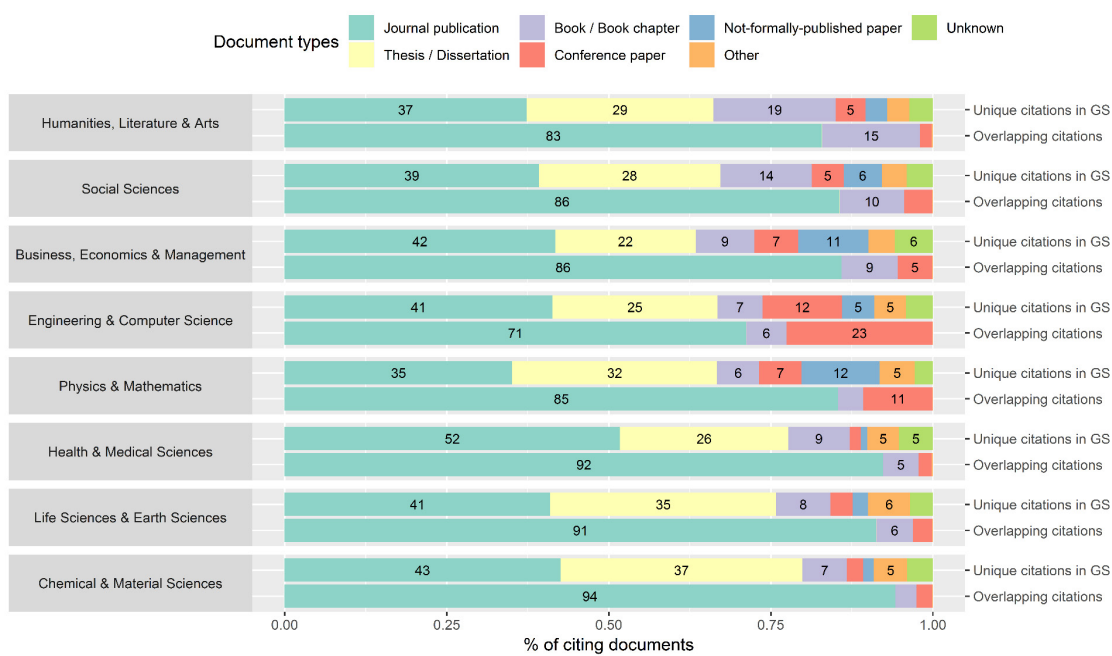

**Figure 6. Distribution of document types among unique and overlapping citations in Google Scholar, by broad subject area of cited documents**

Considering the citations found by WoS and/or Scopus which GS did not find (the citing document might be covered by GS without it making the connection between citing and cited document), most are from journals (Figure 7). Out of the 63,393 citations found by WoS and not by GS (5% of all citations), 41,052 (64% of the WoS citations that GS misses, or 3.2% of all citations analysed in this study) are from journals. Among citations from journal publications, there are more that were published in journals ranked in Q1 and Q4 of their respective JCR categories (0.9% and 1% of all citations), than in Q2 and Q3 (0.6% and 0.5%, respectively). The remaining missing citations come from books or book chapters (19% of WoS citations missing from GS, and 1% of all citations), and conference papers (15% of WoS citations missing from GS, and 0.8% of all citations). The proportions of Scopus citations missing from GS relative to the number of missing citations in GS (136,608) are very similar to those in WoS: 68% of journal publications, 19% books or book chapters, and 13% of conference papers. In this case, the proportion of Scopus citations missing from GS is 9%.

---

[33] https://osf.io/s5ndm/

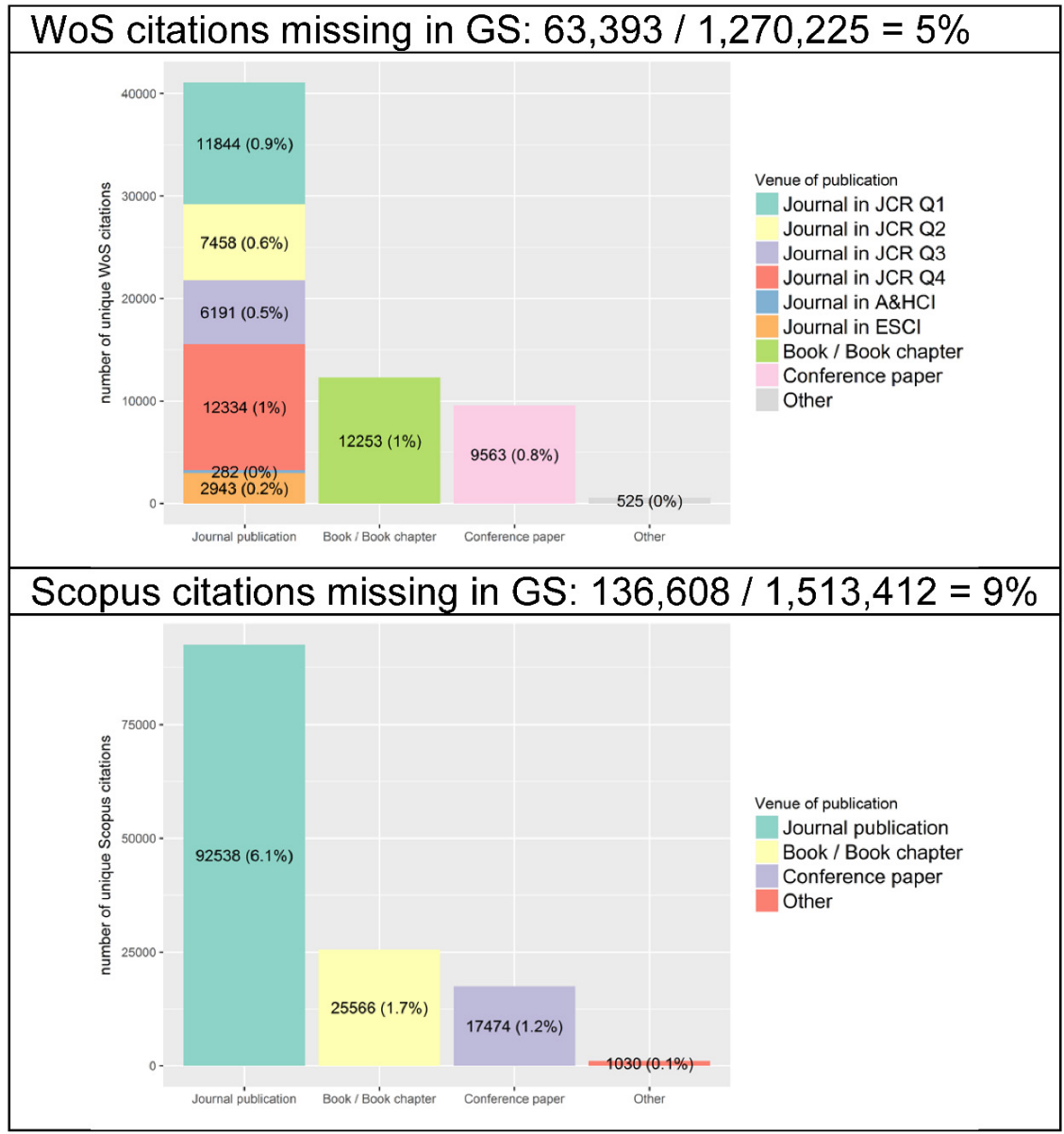

Figure 7. Proportion of document types among citations found by WoS and Scopus, and not by GS

### 3.2.2. Languages

The distribution of languages among the unique GS citations is very different from that of non-unique citations (Figure 8). Whilst for non-unique citations nearly all documents (97%-100%) were published in English, for unique GS citations the percentage ranges from 62% (*Health & Medical Sciences*) to 80% (*Humanities, Literature & Arts*). This is even though all documents in GSCP were published in English. The second most frequent language of unique GS citations was Chinese (4%-12%), and all other languages have a share of 4% or lower across all subject areas. A few (5%-10%) unique GS citations were published in languages outside the top 11 most frequently used languages overall (for all citations in our sample).

At the level of the 252 specific subject categories[34], the categories with a large proportion of non-English unique GS citations are Geochemistry & Mineralogy (59%), Surgery (56%), Radar, Positioning & Navigation (55%), and Cardiology (53%), whereas the categories with the lowest share of non-English citations are Astronomy & Astrophysics (10%), High Energy & Nuclear Physics (11%), Quantum Mechanics (11%), and Computer Hardware Design (11%).

---
[34] https://osf.io/xuz6w/

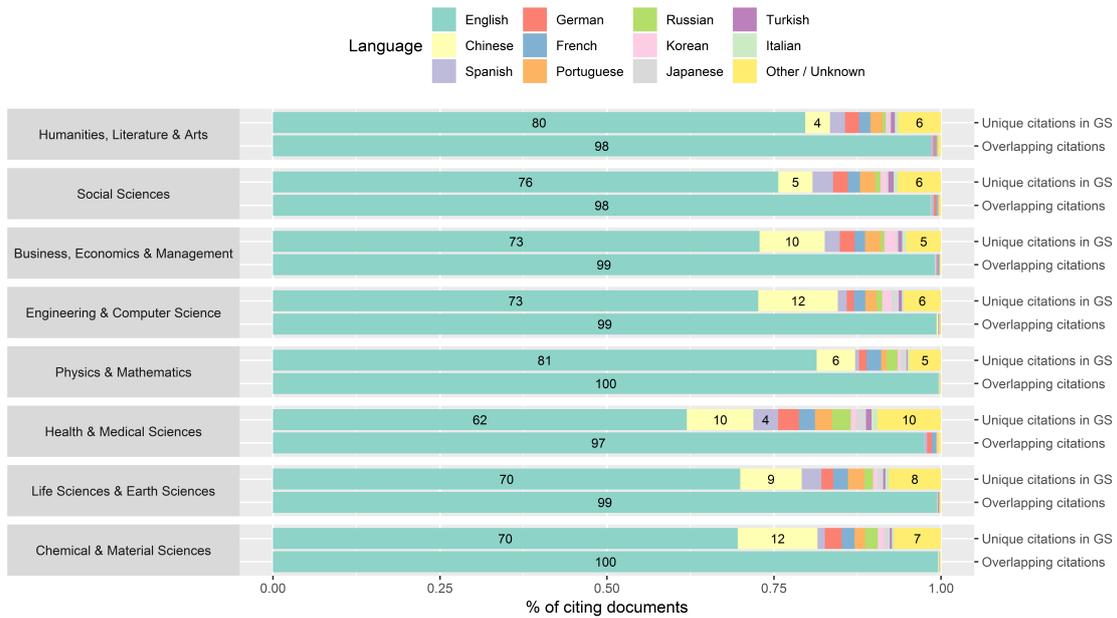

**Figure 8. Distribution of languages among unique and overlapping citations in Google Scholar, by broad subject area of cited documents**

### 3.2.3. Citation counts

This section analyses the Google Scholar citation counts of the 2,301,997 citing documents extracted from Google Scholar. The distributions of log-transformed (ln(1+x) to reduce skewing) citation counts among unique GS citations, and overlapping citations (those also found by WoS and/or Scopus) are different (Figure 9). Across all subject areas, the median log-transformed citation count is always zero and lower than the median of log-transformed citation counts of non-unique citations. The 95% confidence interval for the mean (represented as a red box in Figure 9) is also significantly lower for unique GS citations than for non-unique citations. Both unique and non-unique citations include many outliers (blue dots in Figure 9). The same pattern occurs across the 252 specific subject categories[35], although there are 29 categories in which the median of the citation counts for the unique GS citations is higher than zero (but still lower than the median for overlapping citations).

---

[35] https://osf.io/pm3xh/

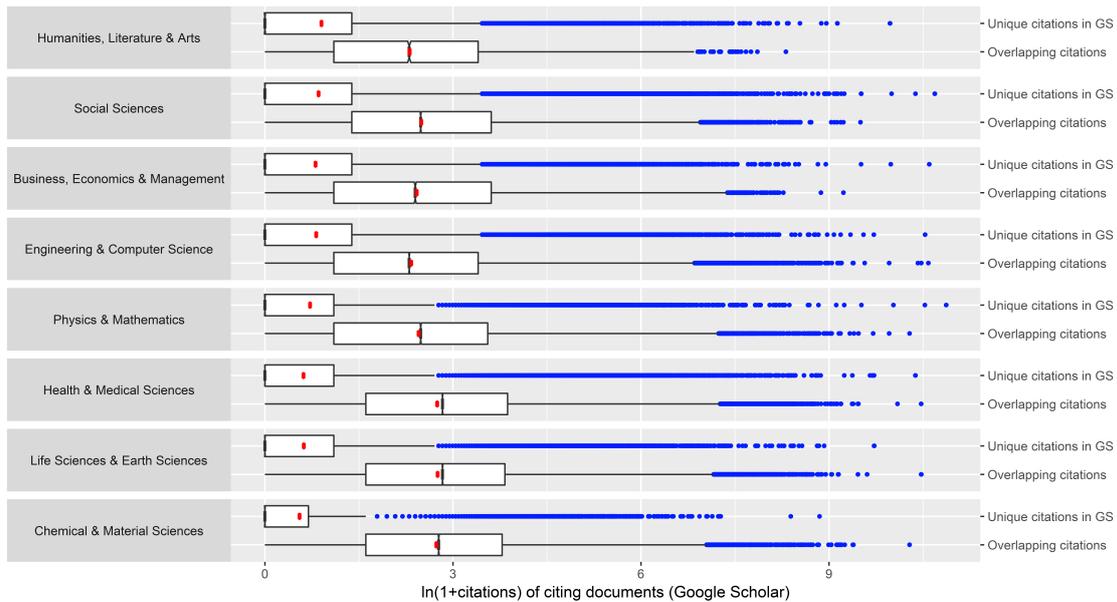

**Figure 9. Distribution of citation counts among unique and overlapping citations in google scholar, by broad subject area of cited document**

## 3.3. RQ3. Citation count comparisons

Spearman correlations between citation counts (GS-WoS, GS-Scopus) are close to 1.0 in most subject categories (Table 3 and Table 4). Correlations between GS and WoS range from .78 in *Literature*, to .98 in *Basic Life Sciences, Biomedical Sciences, Chemistry and Chemical Engineering,* and *Multidisciplinary journals*. In 30 out of the 35 areas of research in the NOWT classification (Tijssen et al., 2010), the Spearman correlation coefficient is over .90. Correlations between Google Scholar and Scopus are even stronger. The weakest correlation is .92 in *Economics, Econometrics, and Finance*, and the strongest is .99 in *Chemical Engineering, Immunology and Microbiology,* and *Multidisciplinary*. In 20 out of 27 categories in the ASJC scheme, correlation coefficients are above .95. The supplementary materials contain tables of citation count correlations computed at the level of the 252 WoS subject categories[36], and the 330 ASJC low-level categories[37], which give broadly comparable results. The weakest statistically significant correlation between GS and WoS at this level[38] is in Medieval & Renaissance Studies (.69), while the weakest correlation between GS and Scopus[39] is .74 in Classics.

On average, GS finds more citations than WoS and Scopus across all categories (see mean citation ratios in Table 3 and Table 4). This effect holds even when citation counts are log-transformed (1+ln(citations)) to reduce skewness. An inverse relationship between strength of correlation coefficients and mean citation ratios of GS over WoS/Scopus is observed. Strong correlation coefficients are associated with lower mean ratios, and vice versa.

---

[36] https://osf.io/x6mw7/

[37] https://osf.io/4pf9z/

[38] https://osf.io/x6mw7/

[39] https://osf.io/4pf9z/

**Table 3. Spearman correlation coefficients, mean ratio, and mean log-transformed citation counts of citing documents between GS and WoS, by subject category**

| Category (NOWT) | N | r | Mean ratio of citation counts GS/WoS | Mean ln(1+citations) GS / WoS |
|---|---|---|---|---|
| Agriculture and Food Science | 24,176 | .97 | 1.74 | |
| Astronomy and Astrophysics | 16,090 | .96 | 1.60 | |
| Basic Life Sciences | 134,045 | **.98** | 1.58 | |
| Basic Medical Sciences | 23,183 | .96 | 1.76 | |
| Biological Sciences | 62,094 | .97 | 1.90 | |
| Biomedical Sciences | 118,817 | **.98** | 1.72 | |
| Chemistry and Chemical Engineering | 129,481 | **.98** | **1.30** | |
| Civil Engineering and Construction | 5,145 | .95 | 1.87 | |
| Clinical Medicine | 223,309 | .97 | 1.82 | |
| Computer Sciences | 61,199 | .86 | 3.21 | |
| Creative Arts, Culture and Music | 1,145 | .84 | 2.80 | |
| Earth Sciences and Technology | 46,536 | .95 | 1.99 | |
| Economics and Business | 28,550 | .93 | 3.30 | |
| Educational Sciences | 13,227 | .92 | 2.92 | |
| Electrical Engineering and Telecommunication | 68,462 | .83 | 3.18 | |
| Energy Science and Technology | 19,242 | .95 | 1.86 | |
| Environmental Sciences and Technology | 64,791 | .97 | 1.86 | |
| General and Industrial Engineering | 10,757 | .92 | 2.37 | |
| Health Sciences | 28,371 | .95 | 2.11 | |
| History, Philosophy and Religion | 5,062 | .90 | 3.15 | |
| Information and Communication Sciences | 6,214 | .94 | 2.87 | |
| Instruments and Instrumentation | 6,167 | .95 | 1.74 | |
| Language and Linguistics | 3,149 | .90 | 3.12 | |
| Law and Criminology | 4,348 | .90 | 3.37 | |
| Literature | 368 | **.78** | **4.02** | |
| Management and Planning | 18,477 | .94 | 2.83 | |
| Mathematics | 17,187 | .91 | 2.65 | |
| Mechanical Engineering and Aerospace | 17,006 | .91 | 2.24 | |
| Multidisciplinary Journals | 44,299 | **.98** | 1.63 | |
| Physics and Materials Science | 144,010 | .97 | 1.52 | |
| Political Science and Public Administration | 8,118 | .90 | 3.22 | |
| Psychology | 32,875 | .95 | 2.50 | |
| Social and Behavioral Sciences, Interdisciplinary | 7,001 | .93 | 2.77 | |
| Sociology and Anthropology | 11,504 | .93 | 2.82 | |
| Statistical Sciences | 12,955 | .92 | 3.16 | |

Confidence level of Spearman correlations: 99%; p-values < 0.01

Highest and lowest values of Spearman correlations and mean citation ratios are highlighted in bold

**Table 4. Spearman correlation coefficients, mean ratio, and mean log-transformed citation counts of citing documents between GS and Scopus, by subject category**

| Category (ASJC) | N | r | Mean ratio of citation counts GS/Scopus | Mean ln(1+citations) GS / Scopus |
|---|---|---|---|---|
| Agricultural and Biological Sciences | 109,423 | .98 | 1.45 | |
| Arts and Humanities | 21,698 | .95 | 2.19 | |
| Biochemistry, Genetics and Molecular Biology | 216,180 | **.99** | 1.43 | |
| Business, Management and Accounting | 40,539 | .94 | 2.43 | |
| Chemical Engineering | 56,569 | **.99** | 1.27 | |
| Chemistry | 118,885 | **.99** | **1.23** | |
| Computer Science | 135,932 | .94 | 1.72 | |
| Decision Sciences | 13,557 | .94 | 2.04 | |
| Dentistry | 3,933 | .97 | 1.78 | |
| Earth and Planetary Sciences | 52,356 | .97 | 1.49 | |
| Economics, Econometrics and Finance | 22,273 | **.93** | **2.83** | |
| Energy | 31,166 | .98 | 1.35 | |
| Engineering | 146,545 | .96 | 1.49 | |
| Environmental Science | 66,212 | .98 | 1.50 | |
| Health Professions | 12,309 | .96 | 1.79 | |
| Immunology and Microbiology | 50,615 | **.99** | 1.44 | |
| Materials Science | 108,794 | .98 | 1.27 | |
| Mathematics | 66,239 | .94 | 1.78 | |
| Medicine | 361,217 | .97 | 1.56 | |
| Multidisciplinary | 18,851 | **.99** | 1.43 | |
| Neuroscience | 46,462 | .98 | 1.55 | |
| Nursing | 19,431 | .96 | 1.80 | |
| Pharmacology, Toxicology and Pharmaceutics | 38,377 | .98 | 1.42 | |
| Physics and Astronomy | 126,820 | .97 | 1.42 | |
| Psychology | 42,037 | .96 | 2.09 | |
| Social Sciences | 81,542 | .94 | 2.22 | |
| Veterinary | 4,550 | .98 | 1.47 | |

Confidence level of Spearman correlations: 99%; p-values < 0.01

Highest and lowest values of Spearman correlations and mean citation ratios are highlighted in bold

# 4. Discussion
## 4.1. Limitations

This study analyses a large sample of citations to highly-cited documents from all subject areas published in English. In order to generalize the results to all articles, it must be assumed that the population of documents that cite highly cited articles is not significantly different from the general population of documents that cite articles. This may not be fully true since, for example, highly cited articles are presumably more likely to be in emerging research areas and larger specialisms. Furthermore, the results may not reflect the citation coverage (in GS, WoS, and Scopus) of documents that do not usually cite scientific literature written in English, such as documents that address locally or regionally relevant topics written in vernacular languages.

Because the highly-cited documents from which our sample of citations came were all initially selected from Google Scholar, this might have provided an advantage to GS in the comparisons: GS might be better suited than WoS or Scopus to find citations for these specific documents, for unknown reasons. Nevertheless, the high citation count correlations found in section 3.3 suggest that this advantage is not substantial, as the three databases provide essentially the same citation rankings at the document level in most subject categories.

Without access to Clarivate Analytics' recently created ESCI Backfile for documents published between 2005 and 2014, an unknown number of citations in this study are listed as found only by GS and/or Scopus, when they are also captured by ESCI. Thus, the results should not be interpreted as applying to all possible WoS data.

Additionally, this article describes a methodology to match citations in GS, WoS, and Scopus at the level of cited articles. The rules chosen to classify a potential match as successful were intentionally conservative to minimize false positives (citations that are matched by the algorithm, despite being different). The matching algorithm probably created some false negatives (citations not matched by the algorithm, despite being the same), especially in categories where DOIs are less widely used and the matching had to rely more frequently on strict title similarity rules. Thus, in some cases the percentages of unique citations might be lower, and percentages of overlaps higher, than reported here.

## 4.2. Comparison with previous studies

The data from previous studies (Table 1) reveal a growth over time in the coverage of citations in GS. While these studies reported that GS could find 38%-94% of all citations found by any source, depending on the discipline(s) of study and the sample analysed, the current study finds values that are higher and more consistent across subject areas. The results here are more similar to those of the most recent study (Moed et al., 2016) and least similar to the earliest studies (Bakkalbasi et al., 2006; Kousha & Thelwall, 2008; Meho & Yang, 2007; Yang & Meho, 2007). For example, GS found 94.3% of all citations to GSCP in *Chemical & Material Sciences*. Although not fully comparable, this figure greatly differs from the 38% of all Chemistry citations found by GS that Kousha & Thelwall (2008) reported. This is evidence that the citation coverage of GS has become much more comprehensive over time. On the other hand, the more recent study by Moed et al., (2016) found that GS contained 94% of all citations in their sample, which is the same as the current study.

The percentages of WoS and Scopus citations that GS could find are generally higher in the current study than previously reported. While prior studies varied greatly depending on the sample (33%-75% of WoS citations, and 44%-89% of Scopus citations), in the current paper GS found 88%-97% of WoS citations, and 84%-94% of Scopus citations (depending on the area). This high relative overlap is a partial cause of the high correlations for citation counts between GS and WoS, and GS and Scopus, found by Martín-Martín et al. (2018). Lastly, this study reports lower percentages of unique citations in WoS (up to 1.9% of all citations) and Scopus (up to 4.3%) than reported in previous studies (up to 23%[40] in WoS, and 12% in Scopus).

---

[40] Considering studies that analysed the three databases (GS, WoS, and Scopus)

Regarding the distribution of document types and languages of GS unique citations, there were substantial percentages of theses and dissertations (from 22% in *Business, Economics & Management*, to 37% in *Chemical & Material Sciences*). These are larger than those found by Kousha & Thelwall (2008), Bar-Ilan (2010), and Lasda Bergman (2012), which found that up to 14% of GS unique citations belonged to this category. In the case of books and book chapters (from 7% in *Chemical & Material Sciences* to 19% in *Humanities, Literature & Arts*), conference proceedings (especially in *Engineering & Computer Science*: 12%), and unpublished materials such as preprints (11% in *Business, Economics & Management*, and 12% in *Physics and Mathematics*), the results are closer to those found by previous studies. The results also show a predominance of English for the citing sources, followed by Chinese (4%-12% depending on the source). These are similar to the results in Kousha & Thelwall (2008) in that Chinese is the second most used language in the sample of citations, although their study found very different percentages (approx. 35% in Biology, 25% in Chemistry, and less than 5% in Physics and Computing).

Lastly, the citation correlations between GS and WoS range from .78 in *Literature*, to .98 in *Basic Life Sciences, Biomedical Sciences, Chemistry and Chemical Engineering,* and *Multidisciplinary journals*, and the correlations between GS and Scopus range from .92 to .99. These correlations are similar to some in previous studies (Amara & Landry, 2012; Delgado López-Cózar et al., 2018; Martín-Martín et al., 2018; Minasny et al., 2013) but somewhat stronger than the ones found by others (De Groote & Raszewski, 2012; Kousha & Thelwall, 2007; Meho & Yang, 2007; Moed et al., 2016; Pauly & Stergiou, 2005; Rahimi & Chandrakumar, 2014; Wildgaard, 2015). This may be due to the disciplines of previous studies or the use of more recent data in the current paper.

## 5. Conclusions

This study provides evidence that GS finds significantly more citations than the WoS Core Collection and Scopus across all subject areas. Nearly all citations found by WoS (95%) and Scopus (92%) were also found by GS, which found a substantial amount of unique citations that were not found by the other databases. In the *Humanities, Literature & Arts*, *Social Sciences*, and *Business, Economics & Management*, unique GS citations surpass 50% of all citations in the area.

About half (48%-65%, depending on the area) of GS unique citations are not from journals but are theses/dissertations, books or book chapters, conference proceedings, unpublished materials (such as preprints), and other document types. These unique citations are primarily written in English, although a significant minority (19%-38% depending on the area) are in other languages. The scientific impact of these unique citations themselves is, on average, much lower than that of citations also found by WoS or Scopus, suggesting that the GS coverage advantage is mostly for low impact documents. Taken together, these results suggest caution if using GS instead of WoS or Scopus for citation evaluations. Without evidence, it cannot be assumed that the higher citation counts of GS are always superior to those of WoS and Scopus, since it is possible that the inclusion of lower quality citing documents reduces the extent to which citation counts reflect scholarly impact. For example, some of the citations from Master's theses may reflect educational impact. Therefore, depending on the type of evaluation that needs to be carried out, it might be necessary to remove certain types of citing documents from the citation counts, as suggested by Prins et al. (2016).

Spearman correlations between GS and WoS, and GS and Scopus citation counts are very strong across all subject categories but weaker in the Humanities (GS-WoS, Literature: .78) and Engineering (GS-WoS, Electrical Engineering and Telecommunication: .83). Also, correlations between GS and WoS (.78 to .98) are weaker than between GS and Scopus (.92 to .99). The weakest correlations are in the categories where there is a greater difference between the citation counts provided by GS, and the citation counts provided by WoS/Scopus. Thus, if GS is used for research evaluations then its data would be unlikely to produce large changes in the results, despite the additional citations found. It would be particularly useful when there is reason to believe that documents not covered by WoS or Scopus are important for an evaluation.

In conclusion, this study gives the first systematic evidence to confirm prior speculation (Harzing, 2013; Martín-Martín et al., 2018; Mingers & Lipitakis, 2010; Prins et al., 2016) that citation data in GS has reached a high level of comprehensiveness, because the gaps of coverage in GS found by the earliest studies that analysed GS data have now been filled. It surpasses WoS and Scopus numerically in all areas of research, and is greatly superior in the areas where WoS and Scopus have a poor coverage, including the Social Sciences and Humanities. However, at this point there is no reliable and scalable method to extract data from GS, and the metadata offered by the platform is still very limited, reducing the practical suitability of this source for large-scale citation analyses, although manual data collection is possible for small scale uses. Nevertheless, providing that a reliable method to extract citation data can be found, the lack of metadata could be solved by combining GS citation data with rich openly accessible data, such as that provided by CrossRef.